# Crystal Structure and Physical Properties of $U_3T_3Sn_4$ (T = Ni, Cu) Single-Crystals

L. Shlyk[1], P. Estrela[2], J.C. Waerenborgh[3], L.E. De Long[4], A. de Visser[2], D.P. Rojas[5], F. Gandra[5] and M. Almeida[3]

[1] *B. Verkin Institute for Low Temperature Physics and Engineering, National Academy of Sciences of Ukraine, 47 Lenin Ave., 310164 Kharkov, Ukraine*

[2] *Van der Waals-Zeeman Institute, University of Amsterdam, Valckenierstraat 65, 1018 XE Amsterdam, The Netherlands*

[3] *Dep. Química, Instituto Tecnólogico e Nuclear, Apartado 21, 2686-953 Sacavém, Portugal*

[4] *Dept. Physics & Astronomy, University of Kentucky, Lexington, KY 40506-0055, USA*

[5] *IFGW, University of Campinas, Campinas, Brazil*

**Abstract**

Heat capacity experiments, crystal structure determination and transmission electron microscopy have been carried out on $U_3Cu_3Sn_4$ single-crystals. $U_3Cu_3Sn_4$ was confirmed to be a heavy-fermion antiferromagnet ($T_N$ = 13(1) K) with a low temperature electronic heat capacity coefficient $\gamma$ = 390 mJ/mol$_U$K$^2$. Low temperature heat capacity experiments on a $U_3Ni_3Sn_4$ single-crystal indicate that below 0.4 K there is a crossover between the previously observed non-Fermi liquid behavior and a Fermi liquid state.

## 1. Introduction

Uranium-based compounds display a variety of magnetic behaviors, ranging from Pauli paramagnetism and spin fluctuations to Kondo effect and long-range magnetic order. The 5f-electron states frequently lie on the itinerant-localised boundary, leading to interesting electronic and magnetic properties. It is generally accepted that the influence of 5f-d or 5f-p hybridization on the physical properties of U-based phases is generally much stronger than that of direct 5f-5f wavefunction overlap. Usually the hybridization strength is quite sensitive to the symmetry of the atomic environment, with higher coordination number corresponding to stronger hybridization. Therefore, systematic studies of the physical properties of different uranium compounds should be performed within isostructural groups.

The ternary $U_3T_3Sn_4$ compounds (where T = Ni, Cu, Pt, Au) crystallize in the cubic $Y_3Au_3Sb_4$-type structure (space group $I\bar{4}3d$) which is a filled variant of $Th_3P_4$-type [1]. The key feature of this system is that the 5f-hybridization of the U atoms can be systematically increased by adding relatively small transition element atoms into voids in the $Th_3P_4$-type structure [1]. Indeed, different magnetic ground states (paramagnetism and weak magnetic order) can be found in the $U_3T_3Sn_4$ family of compounds, including several that exhibit heavy-fermion behavior. Only the Cu variant orders magnetically (antiferromagnet at $T_N = 12$ K) while the other analogues are paramagnetic [2]. Of special interest are the striking departures of the magnetic susceptibility, heat capacity and electrical resistivity from the standard Fermi liquid theory of metals observed in nominally stoichiometric $U_3Ni_3Sn_4$ single-crystals [3], where they were attributed to "non-Fermi liquid" (NFL) behavior.

The growing number of NFL materials has stimulated a number of theories [4-9] that attempt to account for the nonanalytic behavior of several properties at low temperatures, namely the heat capacity, magnetic susceptibility and electrical resistivity. These theories can be grouped into at least two classes: 1) microscopic Hamiltonian models for multichannel or quadrupolar Kondo effects [4], and 2) various types of models that attribute NFL effects to magnetic fluctuations near a zero-temperature quantum critical point (QCP) that separates magnetic order from Fermi liquid (or superconducting) groundstates [5-8].

Recent experimental results [9] indicate the following trends: NFL behavior generally appears on the borderline between magnetic (weakly hybridized) and nonmagnetic (strongly hybridized Fermi liquid) groundstates. There are very few, if any, materials whose physical properties are fully consistent with any single theory for NFL behavior. Most of the relevant systems are alloys, but it is unclear if atomic disorder is a necessary ingredient for NFL behavior. Although it may be difficult to distinguish a unique NFL state from more conventional spin glass ordering, it has been recently recognized that good quality crystals of heavy-fermion compounds such as $UBe_{13}$ and high-$T_c$ superconductors such as $YBa_2Cu_3O_7$, also exhibit similar NFL properties. The latter observations strongly suggest that magnetic frustration or a new type of NFL groundstate is necessary to explain many experimental results, and may require theories based upon a fundamentally new vision of the behavior of matter near absolute zero [10].

The magnetic, transport and specific heat properties of nominally stoichiometric $U_3Ni_3Sn_4$ single-crystals are relatively unique, in that they exhibit NFL behavior at low temperatures [3], that is so far consistent with recent theoretical models based upon classical fluctuations near an antiferromagnetic QCP [8]. Recently, muon spin relaxation experiments (μSR) carried out on the same $U_3Ni_3Sn_4$ single-crystals studied in this work confirmed the absence of magnetic order for $T \geq 2$ K [11]. The $U_3T_3Sn_4$ family is therefore a promising series of compounds to exhibit evidence of an "intrinsic" QCP groundstate in an atomically ordered solid.

In this paper we report the crystal structure determination, TEM analysis, and the magnetic and heat capacity behavior of $U_3Cu_3Sn_4$ single-crystals; and we present supplemental heat capacity data for nominally stoichiometric $U_3Ni_3Sn_4$ single-crystals taken over the extended temperature range 0.1-300 K.

## 2. Single-Crystal Growth and Crystal Structure Determination

The crystals were grown by the Kyropoulos technique from the top of the melt by means of a cooled-seed crystal holder. Bulk charges were first prepared by induction melting with 3:3:4 atomic ratios of pure U (depleted), Ni or Cu, and Sn, respectively, each of at least 99.9% purity. Finally single-crystal samples of $U_3Ni_3Sn_4$ and $U_3Cu_3Sn_4$ were grown by slow cooling a semi-levitated melt of the bulk charges in a cold crucible using an induction furnace. The rate of cooling was approximately 40°C/h between 1600-1400°C and 50°C/h between 1400-800°C. The obtained solid products were about 2-2.5 cm in diameter, and consisted of many crystalline grains, from which single-crystals with dimensions ranging from 0.5 to 2 mm$^3$ were extracted.

Single-crystal X-ray diffraction (XRD) data were collected at room temperature on an Enraf Nonius CAD-4 diffractometer using graphite-monochromated MoK$_\alpha$ radiation ($\lambda = 0.71069$ Å) and an ω-2θ scan mode. The measured intensities were corrected for Lorentz-polarization effects [12] and absorption by an empirical method based on ψ-scans [13]. The crystal structure data for $U_3Ni_3Sn_4$ has been published elsewhere [3]. $U_3Cu_3Sn_4$ data are consistent with the cubic space-group $I\bar{4}3d$. Unit-cell parameters (9.4956(5) Å) were obtained by least-squares refinement of 25 reflections with $11° < 2θ < 45°$. The observed diffracted intensities were refined assuming a $Y_3Au_3Sb_4$-type structure, by a full-matrix, least-squares method [14] based on the squares of the structure factors. The scale factor, an empirical extinction parameter [14], anisotropic temperature factors, and the Sn position parameter were refined. In the last stages of refinement site occupation factors were also allowed to vary. However the best values of the agreement factors (Table 1) were always obtained assuming full site occupancy of the 12$a$ positions by U, the 12$b$ by Cu and the 16$c$ by Sn. A summary of the crystallographic data for the analyzed single-crystals of $U_3Cu_3Sn_4$ is given in Table 1. The interatomic distances and coordination numbers are presented in Table 2.

Specimens for transmission electron microscopy (TEM) were prepared from larger single-crystals of $U_3Cu_3Sn_4$. Flakes of thickness ≈ 120 μm were produced by crushing and placed on a Cu grid, and several electron diffraction patterns were taken over selected areas of a few μm$^2$. Careful examination of several samples revealed no foreign phases or splitting of the diffraction spots due to twinning. As point defects usually have dimensions smaller than the resolution of the electron microscope, we were not able to observe single vacancies. There was no evidence of superlattice reflections due to vacancy ordering both for the $U_3Cu_3Sn_4$ samples. All of the TEM results are similar to those previously reported for an $U_3Ni_3Sn_4$ single crystal [3].

## 3. Results and Discussion

### 3.1. $U_3Cu_3Sn_4$

The specific heat of a $U_3Cu_3Sn_4$ single-crystal was measured in a standard $^4$He cryostat in the temperature range 1.5-30 K using a semi-adiabatic method. The results of the heat capacity measurements for a $U_3Cu_3Sn_4$ single-crystal are shown in Fig. 1. The $C/T$ data show an upturn at low temperatures, and can be extrapolated to a value of 390 mJ/mol$_U$K$^2$ at $T = 0$ K, in agreement with data previously obtained on polycrystalline samples [2]. At $T = 14(1)$ K, the temperature dependence of $C/T$ exhibits a λ-type anomaly typical of long-range antiferromagnetic ordering.

DC magnetization and susceptibility measurements were performed in the temperature range 1.7-300 K in applied fields up to 5 T using a Quantum Design SQUID magnetometer. The magnetic susceptibility of the $U_3Cu_3Sn_4$ single-crystal (Fig. 2) obeys a Curie-Weiss law at temperatures above 30 K with an effective magnetic moment $μ_{eff} = 3.3\ μ_B$ and a Curie-Weiss temperature $θ_p = -11$ K, in good agreement with the polycrystalline results [2] and a maximum in the susceptibility associated with the Néel temperature is present at $T = 12(1)$ K. The low temperature magnetization of $U_3Cu_3Sn_4$ single crystal (see inset in Fig. 2) is linear in field up to 5 T, which is consistent with the antiferromagnetic ground state of this compound. The similarities in behavior between polycrystalline and single-crystal samples of $U_3Cu_3Sn_4$ are important to check, given the significant differences previously reported between poly- [2] and single-crystalline [3] samples of $U_3Ni_3Sn_4$.

It should be mentioned that the heat capacity measurements were done using few single-crystals of $U_3Cu_3Sn_4$ together (the semi-adiabatic method used requires samples of about 0.5 g) while for magnetic measurements only one single-crystal was used. The difference in the Néel temperature obtained from magnetic ($T_N = 12$ K) and thermodynamic ($T_N = 14$ K) measurements as well as the broadening of the λ-type behaviour in the specific heat might indicate a small sample dependence of the antiferromagnetic transition.

### 3.2. $U_3Ni_3Sn_4$

Previously we have described the heat capacity of a $U_3Ni_3Sn_4$ single-crystal [3] in the temperature interval 0.3-5 K using the expression:

$$C = (\gamma_0 - \alpha\sqrt{T})T + \beta T^3 + D/T^2 \quad (1)$$

where $C_E = (\gamma_0 - \alpha\sqrt{T})T$ is the electronic contribution, $C_L = \beta T^3$ the lattice contribution, and $C_N = D/T^2$ represents the high-temperature form of a nuclear Schottky term [15]. The best-fit coefficients obtained were $\gamma_0 = 0.124$ J/mol$_U$K$^2$, $\alpha = 0.0151$ J/mol$_U$K$^{2.5}$, $\beta = 2.07 \times 10^{-3}$ J/mol$_U$K$^4$ and $D = 4.62 \times 10^{-4}$ JK/mol$_U$. From the value of $\beta$ we estimated a Debye temperature $\theta_D \approx 210$ K [3]. This description is in agreement with a renormalization group theory [8] which predicts $\gamma \sim \gamma_0 - \alpha\sqrt{T}$ near a zero-temperature antiferromagnetic instability (NFL behavior). We have estimated a nonuniversal scale factor $T_0 \approx 10$ K using the fitted value of $\alpha = (15/64)k_B N_A N [2/\pi T_0]^{3/2}\xi(5/2)$ [9]. This value of $T_0$ corresponds very well with the onset temperature of the non-analytic behavior of the magnetic susceptibility and resistivity.

Alternatively, recent experimental [16] and theoretical [7] work proposes that NFL behavior might be caused by a competition between RKKY and Kondo interactions in the presence of atomic disorder leading to a Griffiths phase (large magnetic clusters) close to a QCP. We found that the NFL behavior of nominally stoichiometric U$_3$Ni$_3$Sn$_4$ single-crystals can also be described by a divergent power law predicted by this model, i.e., $C(T)/T \propto \chi(T) \propto T^{-1+\lambda}$ with $\lambda = 0.7$ [3]. The best-fit coefficients yielded electronic, lattice and nuclear contributions that differ by only a few percent from those obtained using the renormalization group theory form of Eq. 1.

New heat capacity data spanning the temperature interval 0.1-300 K for a U$_3$Ni$_3$Sn$_4$ single-crystal are presented in Fig. 3. A $^3$He/$^4$He dilution refrigerator was used for attaining temperatures down to 0.1 K, while the high temperature data were obtained in a standard $^4$He cryostat using a semi-adiabatic method. The phononic contribution to the high temperature specific heat can be properly estimated for comparison to previous results. Fitting the high temperature data with a Debye function [17] we obtain the value $\theta_D = 208$ K.

The solid line in Fig.3 is the sum of the numerical solution of the Debye integral and linear electronic contribution.

Data for $C/T$ over the temperature range 0.1-5 K are plotted versus the logarithm of temperature in Fig. 4. The lower temperature data make it clear that the previously calculated nuclear contribution to the specific heat was overestimated [3], since the specific heat data below 0.4 K exhibit a tendency toward saturation, in contrast with what would be expected from an important nuclear contribution to the specific heat, unless a Schottky anomaly (maximum) takes place in the temperature range $0.1 – 0.2$ K. However, this would be very unlikely considering that: i) although quadrupolar interactions from $^{235}$U and $^{62}$Ni could yield a specific heat Schottky anomaly in the measured temperature range, the fact that these isotopes form very small quantities of U$_3$Ni$_3$Sn$_4$ (<1 at.% for the formula unit) makes it difficult to conceive that such contribution would be observable with the relatively large electronic background in U$_3$Ni$_3$Sn$_4$; ii) the $I = 1/2$ Sn isotopes (that have no quadrupolar interactions) would be expected to give rise to a Schottky maximum only at much lower temperatures.

Another possibility would be a crossover to a Fermi liquid regime at the lowest temperatures, which would require a saturation of $C/T$ as $T \to 0$. In fact, the low temperature data below $T = 0.4$ K can be fitted with a modified Fermi liquid expression of the type:

$$C = \gamma T + \delta T^3 \ln(T/T^*) + C_L + C_N \quad (2)$$

where $T^*$ is a characteristic spin fluctuation temperature [18]. The lattice contribution can be fixed using our high temperature Debye fit, while for the nuclear contribution we used $D = 6.45 \times 10^{-8}$ JK/mol$_U$ which corresponds to a typical hyperfine field of $B_{hf} = 0.5$ T for the Sn isotopes with $I = 1/2$, consistent with preliminary Mössbauer experiments on U$_3$Ni$_3$Sn$_4$ single crystals [19]. Using this expression we obtain a characteristic spin fluctuation temperature $T^*$ of the order of 1 K and a Sommerfeld coefficient $\gamma = 0.130$ J/mol$_U$K$^2$. In Fig. 4, we present the fits to the data using the form of Eq. 1 for $T > 0.5$ K, and Eq. 2 for $T < 0.4$ K, considering the same values of $C_L$ and $C_N$ for both temperature ranges. The saturation behavior of the heat capacity can be interpreted as the onset of a degenerate Fermi liquid regime for $T \ll 0.5$ K. Defining $T_{cr}$ as a crossover temperature to a Fermi liquid state, i.e. the temperature below which Eq. 2 applies, we obtain $T_{cr} = 0.40(2)$ K. With this reduced nuclear contribution Eq. 1 only holds down to $T \sim 0.5$ K, which means that there is a crossover region in the temperature range 0.4-0.5 K separating a non-Fermi liquid to a Fermi liquid regime.

An assessment of the potential existence of a Schottky term in the specific heat can only be made by studying the effect of a magnetic field, or the acquisition of data for $T < 0.1$ K.

## 4. Conclusion

We have characterized the heat capacity, microstructure and crystal structure of $U_3Ni_3Sn_4$ and $U_3Cu_3Sn_4$ single-crystals. $U_3Cu_3Sn_4$ is a heavy-fermion antiferromagnet while nominally stoichiometric $U_3Ni_3Sn_4$ is a nearly magnetic NFL compound.

We find that satisfactory fits of the heat capacity data for $U_3Ni_3Sn_4$ in the temperature range 0.5-5 K always require a dominant electronic term which exhibits a near-square-root temperature dependence, consistent with a theoretical model for NFL systems near a zero-temperature quantum transition from magnetic to non-magnetic states [8] or an alternative Griffiths phase model [7]. It should be noted that the renormalization group theory treatment does not include the effects of disorder, which must be present in real systems to some degree, while the Griffiths phase model includes disorder as a crucial ingredient.

Previously we have undertaken a thorough analysis of the heat capacity and susceptibility of nominally stoichiometric $U_3Ni_3Sn_4$ [3], and conclude that several NFL models (e.g., multichannel Kondo [4] and Kondo disorder [20]) commonly considered in the contemporary literature do not describe the entire data set known for this material. All attempts to include a logarithmic heat capacity term resulted in an unphysically high characteristic temperature scale $T_0$ [3]. On the other hand, the scaling temperature $T_0 \approx 10$ K, obtained from low temperature analysis based on renormalization group theory, is consistent with the onset of the non-analytic behavior of physical properties in this material. Having extended heat capacity measurements to the lower temperatures we found a crossover to a Fermi liquid state below 0.4 K described by a $T^3 \ln T$ term to the specific heat, characteristic of spin fluctuation phenomena. The crossover to a Fermi liquid state is a characteristic of NFL materials that are imprecisely tuned to a QCP [8], due to unfavourable conditions of a "control parameter" such as pressure, magnetic field, composition or atomic disorder. In the case of nominally stoichiometric $U_3Ni_3Sn_4$, such a non-thermal critical parameter might be vacancy doping which governs the degree of spd-f hybridization leading to competition between NFL-FL states. On the other hand, a possible marginal Fermi liquid ground state can not be ruled out. The precise role of small amounts of disorder detected by XRD analysis in nominally stoichiometric $U_3Ni_3Sn_4$ single-crystals must be investigated by further studies of carefully characterized samples.

The moderate size of the Sommerfeld coefficient $\gamma \sim 0.130$ J/mol$_U$K$^2$ and reduction of $\mu_{eff} = 2.0$ $\mu_B$/U [3] from 3.62 or 3.58 $\mu_B$ expected for U$^{3+}$ or U$^{4+}$ free ions, respectively suggest significant hybridization between 5f and itinerant electron states takes place in $U_3Ni_3Sn_4$. The replacement of the Ni atoms by Cu gives rise to an increase in the number of 3d electrons, leading to a dehybridization of the 5f states with the 3d band. This is consistent with the observed lattice expansion on replacing Ni (9.3577(4) Å) with Cu (9.4956(5) Å), the increase of $\gamma$ by a factor of about 3, and development of antiferromagnetism below 14 K in $U_3Cu_3Sn_4$ with an effective magnetic moment $\mu_{eff} = 3.3$ $\mu_B$/U, corresponding to either a 5f$^2$ or 5f$^3$ uranium configuration. The apparent variation in 5f hybridization with transition element substitution, the crossover from a Fermi liquid ground state in $U_3Ni_3Sn_4$ to weak antiferromagnetism with $T_N = 14$ K in $U_3Cu_3Sn_4$, and the observation of NFL properties in $U_3Ni_3Sn_4$ for the temperature range 0.5 – 5 K, imply that a QCP should exist in this series of materials over some range of composition and/or pressure in the vicinity of stoichiometric $U_3Ni_3Sn_4$.

**Acknowledgments**

The single-crystals were grown by L.S. during her stay in Portugal under the support of a NATO fellowship. Research at the University of Kentucky was supported by NSF Grant #INT-9515504. P.E. acknowledges the European Commission for a grant within the TMR Programme. Part of this work was supported by FAPESP (Brazil).

Table 1 - Crystal data and details of crystal structure refinement for $U_3Cu_3Sn_4$.

| | |
|---|---|
| Ideal chemical formula | $U_3Cu_3Sn_4$ |
| Formula weight | 1379.5 g mol$^{-1}$ |
| Crystal system | Cubic, Body-centered |
| Space group | $I\bar{4}3d$, (No.220) |
| $a_0$(300 K) | 9.4956(5) Å |
| V | 856.18 Å$^3$ |
| Z | 4 |
| $\mu$(MoK$_\alpha$) | 75.16 mm$^{-1}$ |
| Approximate crystal dimensions | 0.07mm x 0.07mm x 0.37mm |
| Radiation, wavelength | MoK$_\alpha$, $\lambda$=0.71069 Å |
| Monochromator | Graphite |
| Temperature | 295 K |
| $2\theta$ range | 2°-80° |
| $\omega$-2 scan$\theta$ | $\Delta\omega$=0.90+0.35 tan $\theta$ |
| Data set | -17$\leq$ h $\leq$17, 0$\leq$ k $\leq$17, 0$\leq$ l $\leq$17 |
| Crystal-to-receiving-aperture distance | 173 mm |
| Horizontal, vertical aperture | 4mm, 4mm |
| Total data | 2648 |
| Unique data | 450 |
| Observed data (Fo $\geq$ 4$\sigma$(Fo)), n | 430 |
| Number of refined parameters, p | 9 |
| Final agreement factors | |
| $R1 = \sum ||F_o|-|F_c|| / \sum |F_o|$ | 0.0226 |
| $wR2 = \left[\sum\left[w\left(F_o^2-F_c^2\right)^2\right]/\sum\left[w\left(F_o^2\right)^2\right]\right]^{1/2}$ | 0.0667 |
| $GooF = \left[\sum\left[w\left(F_o^2-F_c^2\right)^2\right]/(n-p)\right]^{1/2}$ | 0.506 |

Table 2 - Interatomic distances $d$ (up to 5 Å) and average number $NN$ of nearest neighbours of U, Cu and Sn atoms in $U_3Cu_3Sn_4$.

| | NN | Atoms | d | | NN | Atoms | D |
|---|---|---|---|---|---|---|---|
| U(12a) | 4 | Cu | 2.907 | Sn(16c) | 3 | Cu | 2.654 |
| | 4 | Sn | 3.287 | | 3 | U | 3.287 |
| | 4 | Sn | 3.286 | | 3 | U | 3.286 |
| | 8 | U | 4.441 | | 3 | Sn | 3.538 |
| | 2 | Cu | 4.748 | | 2 | Sn | 4.112 |
| | | | | | 3 | Cu | 4.279 |
| Cu(12b) | 4 | Sn | 2.654 | | 6 | Sn | 4.682 |
| | 4 | U | 2.907 | | | | |
| | 4 | Sn | 4.279 | | | | |
| | 8 | Cu | 4.441 | | | | |
| | 2 | U | 4.748 | | | | |

**Figure captions**

Fig. 1 - Temperature dependence of the specific heat divided by temperature for single crystalline $U_3Cu_3Sn_4$.

Fig. 2 - The magnetic susceptibility $\chi$ and $\chi^{-1}$ as a function of temperature for a $U_3Cu_3Sn_4$ single crystal. The solid line is a fit to a Curie-Weiss law. The inset shows the magnetization vs. magnetic field at $T = 1.7$ K.

Fig. 3 - Temperature dependence of the specific heat of single-crystalline $U_3Ni_3Sn_4$. The solid line is a fit to data with the sum of the Debye approximation of the phonon contribution and linear electronic term to the specific heat.

Fig. 4 - $U_3Ni_3Sn_4$ low temperature specific heat divided by temperature versus $\ln T$. The lines are fits to the data with Eqs. 1 and 2 (see text).

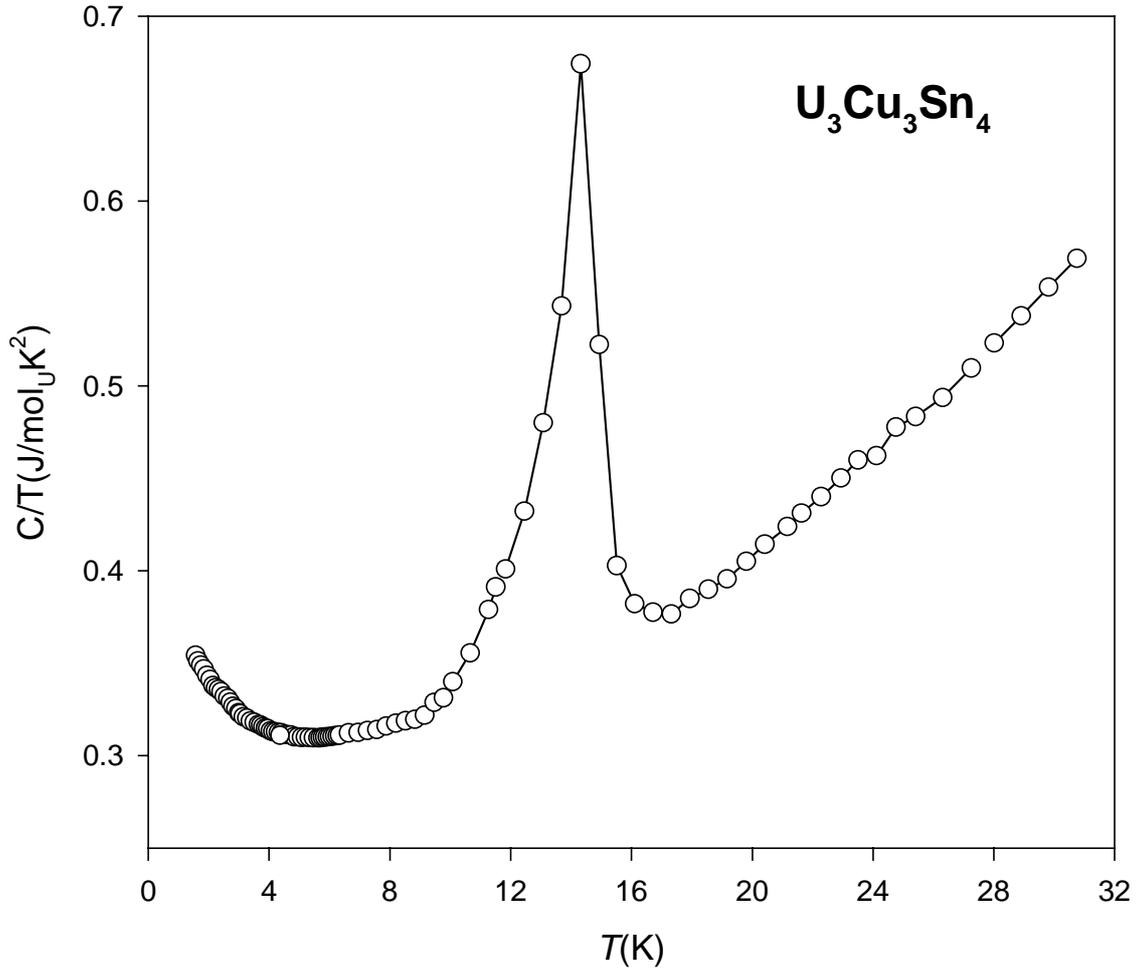

Fig. 1 - L. Shlyk et al.

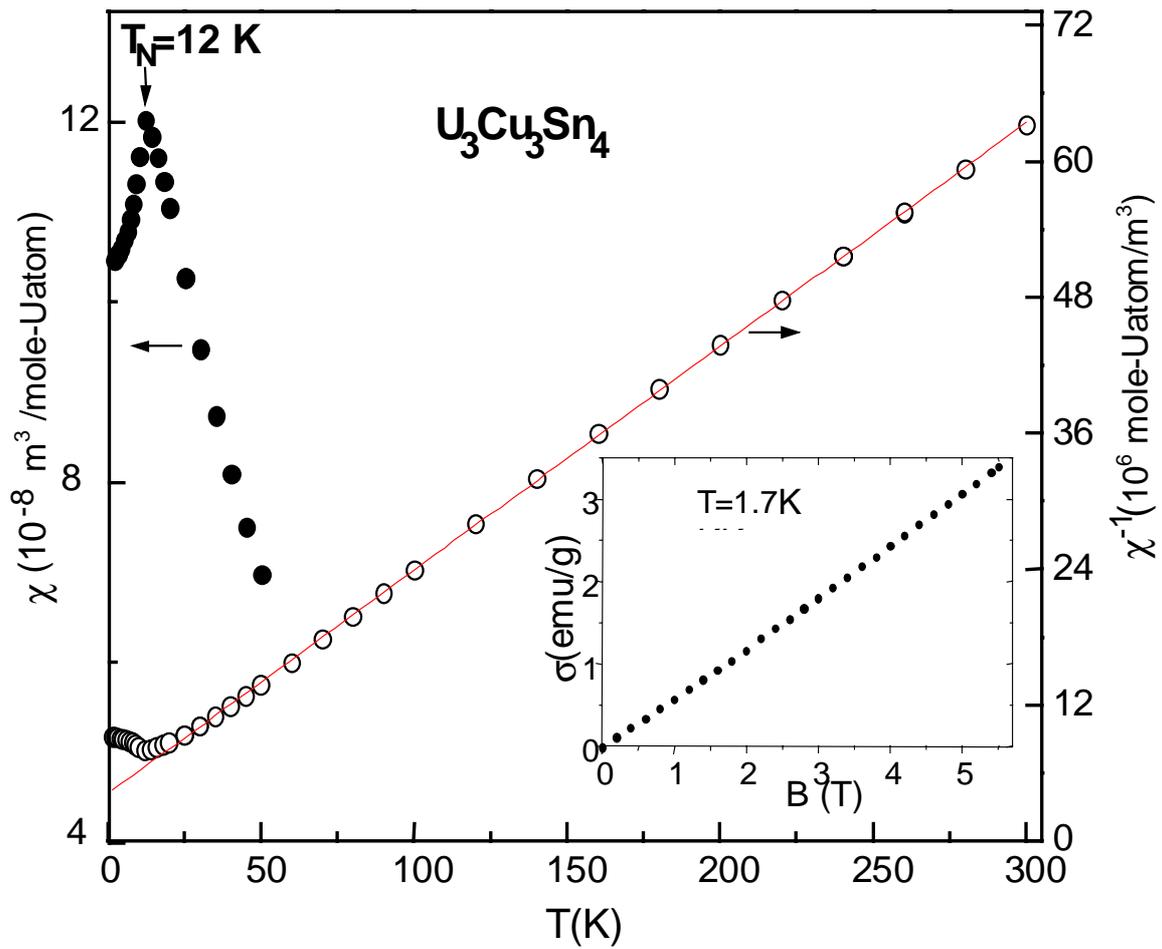

Fig. 2. - L.Shlyk et al.

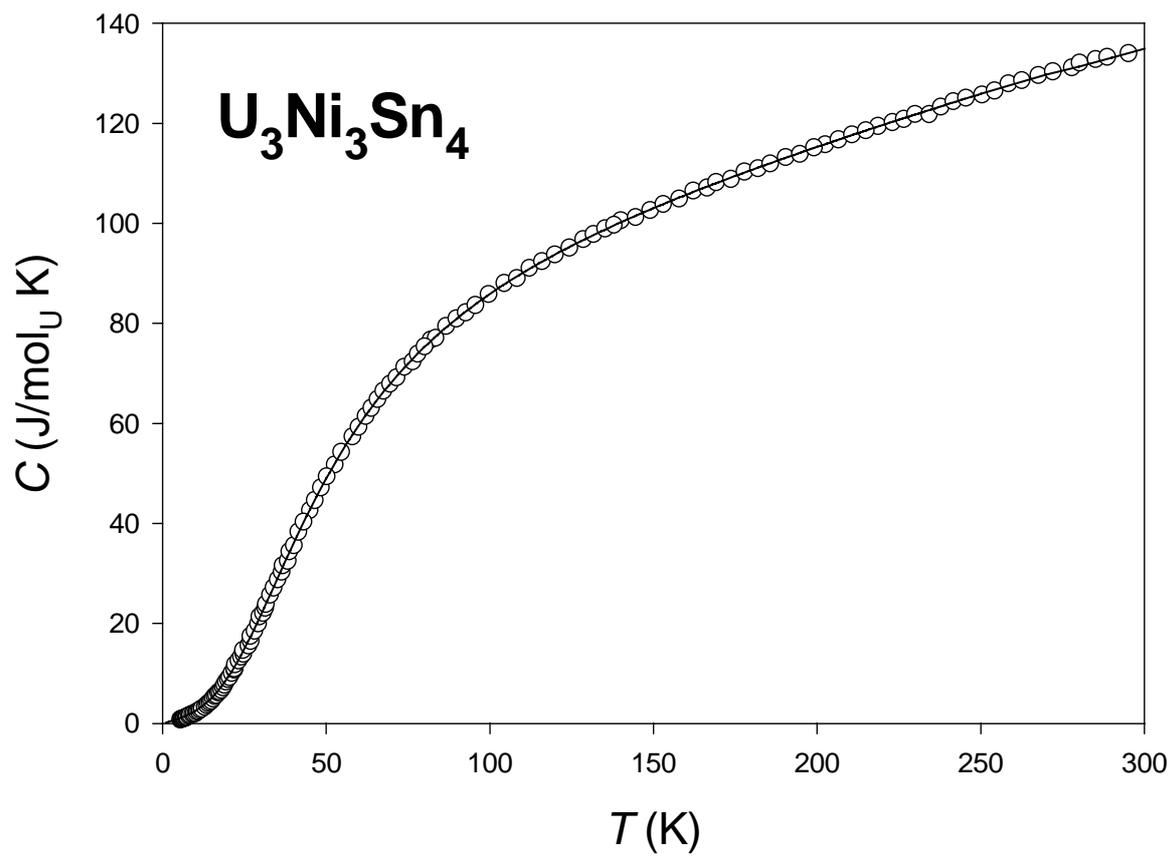

Fig. 3 - L. Shlyk et al.

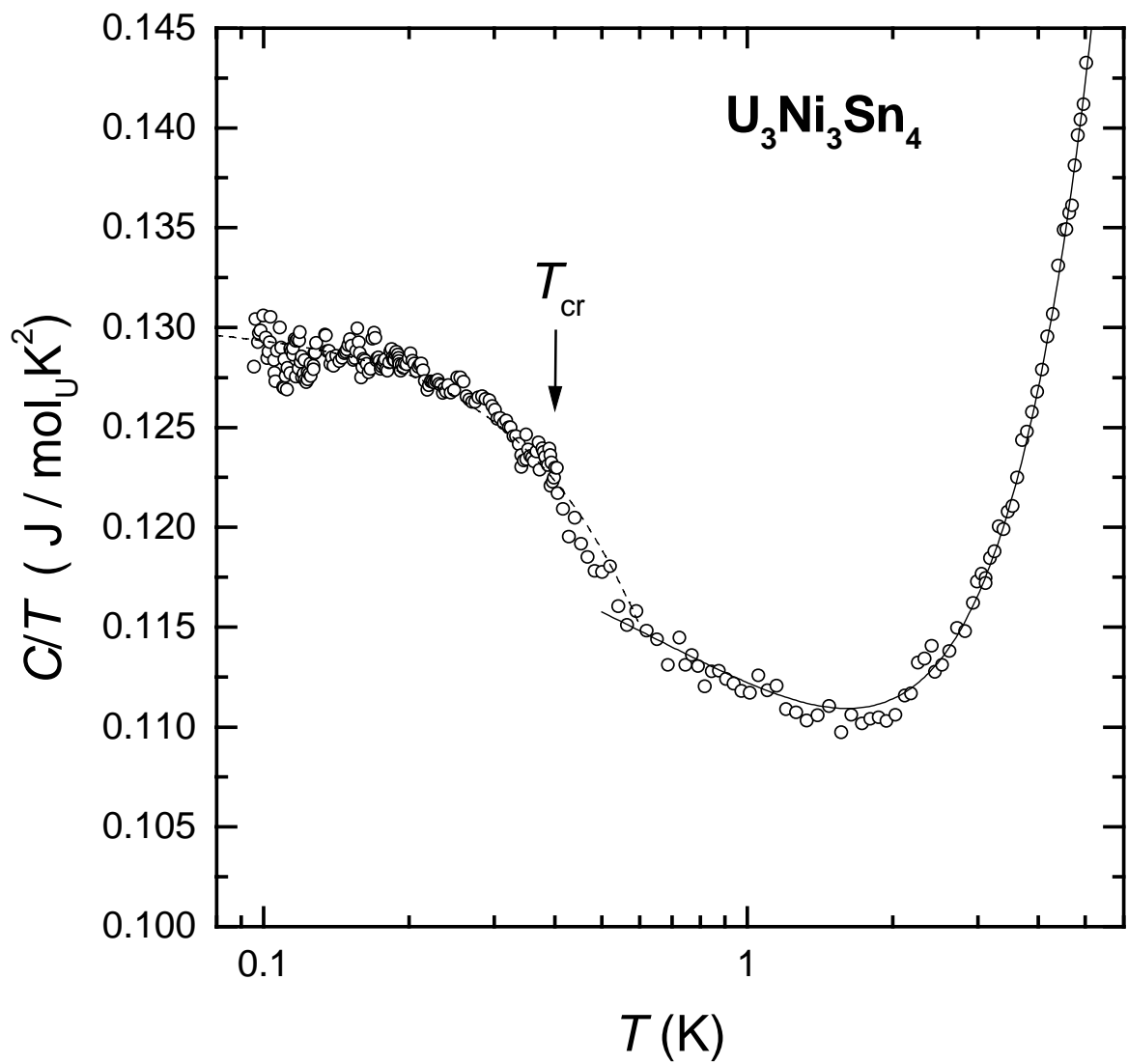

Fig. 4 - L. Shlyk et al.